\documentclass[11pt,twoside]{article}
\usepackage{asp2010}

\resetcounters

\begin{document}

\title{GALFACTS:  The G-ALFA Continuum Transit Survey}
\author{A.\ R.\ Taylor$^1$ and C.\ J.\ Salter$^2$
\affil{$^1$Department of Physics and Astronomy, University of Calgary, \\
2500 University Dr. N.W., 
Calgary, Alberta, Canada, T2N 1N4 \\
$^2$ National Astronomy and Ionosphere Center, Arecibo Observatory \\
Arecibo, Puerto Rico 00612, USA}}

\begin{abstract}
The GALFACTS project is using the L-band seven feed array receiver system on the
Arecibo telescope to carry out an imaging spectro-polarimetric survey of the 30\% of the sky visible from Arecibo. 
GALFACTS observations will create full-Stokes image cubes at an
angular resolution of 3.5$'$, with several thousand spectral channels covering 1225 - 1525 MHz, and band-averaged sensitivity of 90\,$\mu$Jy, allowing sensitive imaging of 
polarized radiation and Faraday Rotation Measure from both diffuse emission and against a high density grid of extragalactic sources. GALFACTS will be a major observational advance in imaging of the polarized radiation from the Milky Way and will provide a rich new database for exploration of the magnetic field of the Galaxy and the properties of the magneto-ionic medium. 
\end{abstract}

\section{Introduction}
The Arecibo 305-m antenna is the world's largest single-dish telescope, achieving
continuous spatial frequency coverage, high brightness sensitivity,
and arcminute-scale angular resolution at decimetre wavelengths.  The new ALFA
multi-beam receiver system allows these properties to be
used to efficiently image large areas of the sky at $\lambda$21\,cm.  
The GALFACTS Consortium is using the Arecibo telescope and ALFA 
to carry out a sensitive, high resolution, spectro-polarimetric 
survey of the region of the sky visible with the Arecibo telescope -- 
the GALFA Continuum Transit Survey, GALFACTS.

A key observational objective of the GALFACTS  is to image the  
polarized emission from both discrete objects and the diffuse
interstellar medium of our Galaxy and to derive polarization properties,
including Faraday Rotation Measures for a vast population of
extragalactic sources.
Wide-field polarimetric imaging of Galactic emission has enjoyed a surge of activity
over the past decade with the detection of highly-structured,
arcminute-scale polarized radiation in interferometric images, 
both at high Galactic latitude, e.g.  in WSRT 349-MHz images \citep{Haverkorn_2003},
and in the plane of the Galaxy with the Canadian Galactic Plane Survey 
\citep{CGPS, CGPS_pol} and Southern Galactic Plane Survey
\citep{SGPS_pol} at 1.4 GHz.  Similar structures are seen at 
5 GHz by \cite{Sun_2007}.
These structures are superposed on the polarized emission from
the diffuse Galactic synchrotron emission, but themselves have no
Stokes-I counterpart.
The accepted interpretation is that the
distributed polarized emission arises from the intrinsically-smooth
Galactic synchrotron emission, but differential Faraday rotation in the
intervening magneto-ionic medium (the Faraday Screen) imposes fine
structure due to the resulting spatial variation in polarization position angle.
Propagation effects in the interstellar medium dominate over intrinsic polarized 
emission structure.  
Observations of diffuse polarized emission and of the propagation effects of the
ISM on the Rotation Measures of background sources are thus 
powerful probes of both the relativistic and thermal plasmas of the 
interstellar medium.

GALFACTS promises a major observational advance in this field by 
virtue of its high brightness sensitivity compared to interferometric
surveys and the high angular resolution compared to other
single-dish telescopes as well as the large spectral bandwidth combined 
with high spectral resolution that will allow accurate measurement
of the frequency dependence of the polarization state of the emission at
1.4 GHz.
Presently, the highest resolution, single-dish, L-band continuum surveys 
are those of the northern sky by the Effelsberg telescope
\citep{Reich_1997} and in polarization by \citep{Uyaniker_1999}
of a few selected regions at intermediate latitudes, both with 
resolution of $9.4^{\prime}$ and brightness sensitivities of 10's of mK.
For surveys covering a significant fraction of the sky in polarization, 
only significantly poorer resolution data are available; 
$\theta_{\rm{HPBW}} = 35^{\prime}$,  
\citep{Wolleben_2006} at similar sensitivity (12 mk).
In comparison, GALFACTS will provide full-Stokes image cubes at 
resolution of 3.5$^{\prime}$, at a brightness sensitivity an order 
of magnitude deeper than previous work, and covering a frequency band from
1225 -- 1525\,MHz in spectral-line mode, allowing imaging of 
Faraday Rotation Measure.

\section{Observations}

\begin{figure}[!ht]
\plotfiddle{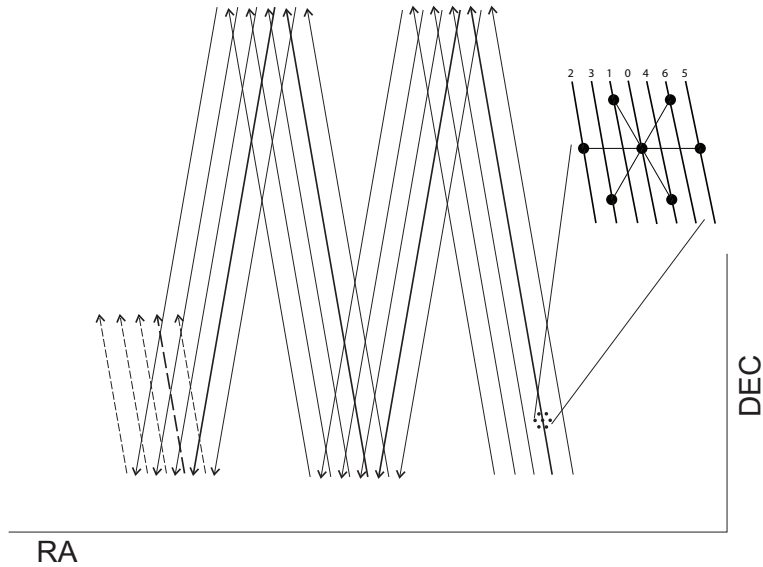}{2.9in}{0}{62}{62}{-135}{0}
\label{Taylor_fig1}
\caption{GALFACTS observations are made by scanning the seven-feed ALFA receiver up and down the 
meridian at a rate of 1.53 degrees per sidereal minute.  Each day of observations creates
a zig-zag track of up and down ``scans'' each consisting of seven sub-tracks (see inset
at upper right), one from each of the ALFA beams. Main tracks for successive days of observations 
are separate by 51 seconds of Right Ascension. }
\end{figure}

The observations are carried out with the Arecibo L-band feed array system, ALFA.  
The seven feed horns of the ALFA provide seven beams on the sky with six beams arranged
in a hexagonal pattern around a central on-axis beam.   Each feed horn detects two 
nominally orthogonal states of linear polarization. 
Data is acquired simultaneously in each polarization channel while the feed system is 
alternately scanned north and south along the meridian at a rate of 1.53 degrees per sidereal minute.
Combined with the Earth's rotation, this tracking speed creates a zig-zag track pattern
on the sky with the tracks from individual beams separated by 1.83$'$, close to
Nyquist sampling for the 3.5$'$ FWHM beams.  
The tracking geometry is illustrated in Figure~\ref{Taylor_fig1}. 
On consecutive days the track of the central beam is shifted by 51 seconds of Right Ascension. 
For 18 degree long Declination scan, 28 separate days/tracks provides complete coverage.

\begin{figure}[!ht]
\plotone{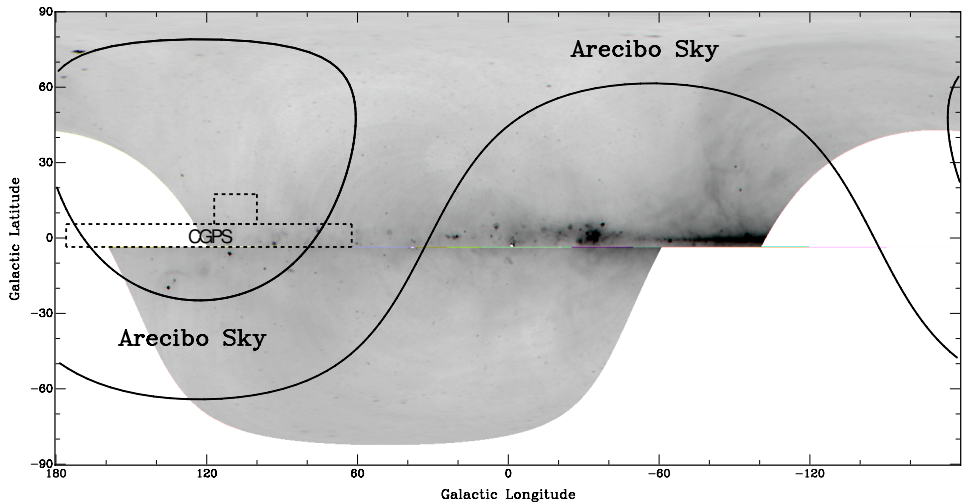}
\label{Taylor_fig2}
\caption{The 12,734 square degrees to be covered by GALFACTS is shown superposed on the
Effelsberg 1420 MHz continuum image of the northern sky.  GALFACTS crosses the
Galactic plane in the first and third quadrants, complementing the coverage of the Canadian
Galactic Plane Survey (shown by the dashed lines). It will also cover large regions at mid latitudes
in the south and will extend from the plane to the pole over two ranges of longitude in the inner 
and outer Galaxy.}
\end{figure}

GALFACTS observing is separated into three survey regions; 
a south survey consisting of the portion of the sky south
of zenith angle 1.5$^{\circ}$ on the southern meridian, 
($-0.8^{\circ} < \delta < 16.8^{\circ}$), a north survey of the sky north
of zenith angle 1.5$^{\circ}$ on the northern meridian 
($19.8^{\circ} < \delta < 37.8^{\circ}$), and a zenith survey of the
strip within $Za < 2.0^{\circ}$, 
($16.3^{\circ} < \delta < 20.3^{\circ}$).  Together these provide
complete coverage of the 12,734 square degrees of sky 
from ($-0.8^{\circ} < \delta < 37.8^{\circ}$).  
Figure~\ref{Taylor_fig2} shows the survey coverage in Galactic coordinates superposed on the
Effelsberg 1.4 GHz continuum survey of the northern sky \citep{Reich_1997}.
Observations are carried out at night to avoid the effects of solar radiation
in the sidelobes of the antenna which can produce large spurious polarized signals.  
Observations are scheduled in 6-hour RA blocks,with 12 blocks and a total observing time of 
1600 hours required for the entire survey.
For each observing block a series of calibration observations of strong compact sources are
also obtained measuring the full-Stokes response of the seven ALFA beams as a function
of frequency and elevation.
The first observing block occurred in December 2008 and observations are scheduled to be completed
by 2013.  

\section{Data Acquisition and Processing}

The FPGA spectrometer backend (the Mock spectrometers) produces
data streams for each of four polarization states and 
4096 spectral channels for all seven beams in two adjacent 172-MHz bands, which
together cover the 300 MHz total band from 1225-1525 MHz.
During the observations a continuous square-wave noise diode signal is injected into the two 
polarization channels at the feed horns at 25 Hz with 50\% duty cycle
to calibrate the time dependence of the complex gains of the receiver system.  
The noise diode is injected with equal strengths into the X and Y polarization channels.
Each of the fifty-six (seven beams, four polarizations, two bands) 4096-channel data streams 
are sampled at 1 millisecond per sample to allow the on and off states of the noise diode to be 
detected in post processing.   The aggregate data rate is 460 MB/s. 

Data are recorded directly onto disk drives at Arecibo Observatory.  After each observing day the data are
run through first-stage processing which separates the on and off states of the noise diode signal and time averages the data to 0.2 s sample time diode-on and diode-off data streams.  These time-averaged data are 
shipped to the University of Calgary at the end of each observing run for further processing.  

Calibration
of the data is simplified by the fact that we observe only along the meridian at a fixed feed rotation.  There
are thus only three independent variables for instrumental effects, frequency, time and elevation angle.
The data from each day and beam are corrected for time variations of the receiver channel 
gains using the noise diode signals. This process converts the raw data into Stokes parameters in units 
of brightness temperature.  Frequency-dependent
baselines are remove by fitting polynomials to the elevation dependence of the signal amplitude in regions
removed from strong sources for each day, beam and frequency channel.  
The Stokes $I$ band shape and the polarization leakage derived from calibration observations are then applied
and the corrected data are ``basketweaved'' and gridded into images.  
The frequency-dependent baseline subtraction removes the emission on spatial scales of order
10 degrees and longer. These missing spatial frequencies will be recovered by integrating
data from the Global Magneto-ion Medium Survey, GMIMS \citep{Wolleben_2009}.  Further processing,
including cleaning the dirty images to remove sidelobes from the multiple beams 
 \citep{guram_2008} is carried out on the gridded image data.

The final data products from GALFACTS are spectro-polarimetric image cubes, 
(intensity as a function of $\alpha$, $\delta$, $\nu$), in each Stokes parameter I, Q, U and V,
with $\sim$8000 spectral channels over a 300 MHz bandwidth.  
For a 3.5$'$ FWHM beamsize the effective integration time per scan for a given direction on the
sky is 2.4s, or a total integration time per beam area of 4.8s.
For an average SEFD over the seven beams of 3.4 Jy, the theoretical noise level 
per 42 kHz channel is 7.5 mJy per polarization.
Integrated over the full 300-MHz band the theoretical $1\sigma$ point-source sensitivity 
level is $\sim90$ $\mu$Jy per polarization, or a surface brightness sensitivity of 1 mK.  

\begin{figure}[!ht]
\plotfiddle{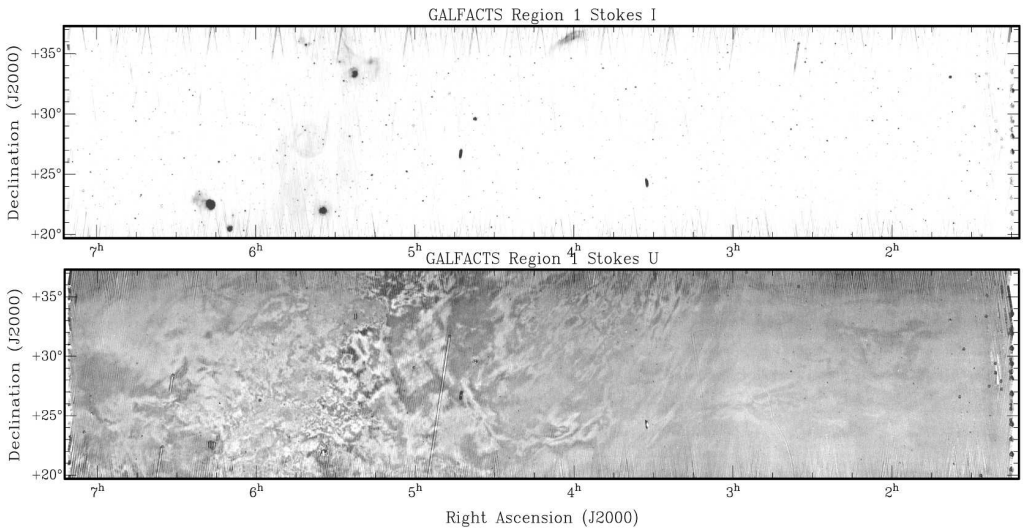}{3in}{0}{96}{96}{-205}{0}
\label{Taylor_fig3}
\caption{Preliminary Stokes I (top) and Stokes U (bottom) images of the first GALFACTS observing
block covering 18 degrees of declination over the RA range 01$^{\rm h}$ 14$^{\rm m}$ to
7$^{\rm h}$ 10$^{\rm m}$. 
These images, which are integrated over 10 frequency channels, have a bandwidth of 0.4 MHz.
The structures seen in $U$ are primarily produced by spatial variations in
polarization position angle from Faraday rotation in the Galactic magnetic-ionic medium. }
\end{figure}

A preliminary image from the first GALFACTS observing run is shown in Figure~\ref{Taylor_fig3}.
These images show emission over a 0.4 MHz bandwidth at 1473 MHz. The Stoke $U$ image
shows emission throughout, primarily due to spatial variations in polarization
position angle due to the Faraday screen of the Galactic magneto-ionic medium.  The band
of highly-structured emission between 5$^{\rm h}$ and 6$^{\rm h}$ RA runs along the Galactic 
plane. As one moves above and below the plane the polarization structures persist but become
smoother and larger scale.  

\section{GALFACTS Science Goals}

The GALFACTS project is being carried out by a consortium of researchers from Canada, Australia, Europe, India 
and the United States.  GALFACTS will provide a legacy data set that will be a valuable complement to other 
large-scale surveys of the Galactic ISM and the extragalactic sky - in particular surveys that will be carried
out in the post 2013 era by Square Kilometre Array precursor telescopes.  In addition the Consortium has identified a list of immediate science goals for the project, including:

\begin{itemize}
\item Imaging of the polarized emission from our Galaxy from the plane to the pole, permitting a study 
of the arcminute-scale polarization seen at all latitudes for frequencies between 330 MHz and 
1.4 GHz and their relation to the ISM.  Such structures appear ubiquitously and on all observed spatial scales
and reflect processes at pc to kpc scales in the magneto-ionic medium. 

\item Measuring the polarization properties of a large sample of compact extragalactic radio sources down 
to polarized
flux density levels of a few 100 $\mu$Jy.  The polarization sensitivity of GALFACTS is 
several times deeper than the NVSS, and will provide 
a statistical probe the polarization of the faint radio source population, which has been recently shown by \cite{Taylor_2007} and \cite{Grant_2010} to be more highly polarized than previously thought.  GALFACTS will allow the measurement of the polarization percentages, position angles and rotation measures of an estimated 
100,000 extragalactic radio sources. 
While powerful AGNs still dominate the source population at this level, some 25\% of the sources are likely to be normal or starburst galaxies, low luminosity AGNs (e.g. LINERS), and Seyfert galaxies.

\item Investigate the role of magnetic fields in the cold neutral medium, and especially the contribution of these fields in the formation of molecular clouds and in generating the conditions for star formation.

\item Studies of discrete Galactic radio sources such as supernova remnants (SNR) and HII regions. GALFACTS Stokes I images are expected to uncover new, low surface brightness examples of both these classes of objects, while the parallel polarization information will provide a wealth of information on both newly discovered objects and already known examples. For example, the Faraday depolarization of the diffuse background radiation by HII regions is a unique way to study the magnetic fields within these objects. In the case of SNRs, the GALFACTS images will allow us not only to study the orientation of magnetic fields in the shells of these objects, but also to investigate the strength of these fields and their relation to the large scale galactic
magnetic field, see, for example, \cite{Kothes_2009}.

\item Study the global structure of the Galactic magnetic field, investigating both the large-scale ordered field, and the turbulent small-scale component. This will be achieved through both the diffuse Galactic
polarized emission and the catalog of extragalactic RMs, which will contain a rich harvest of information on the Galactic magnetic field, see, for example, \cite{Brown_2007}.  Identification of field reversals in the spiral magnetic pattern of the Galaxy and determination of the polarity of its vertical structure are key to understanding the process that has produced the field  

\item GALFACTS sensitivity is sufficient to detect the effects of magnetic fields in the halo of our Galaxy, allowing investigation of the properties of turbulence within the ionized halo gas, and studying the disk-halo interface.  The structure of the field in the interface region is unknown, and yet it is probable that magnetic fields play an important role in the exchange of mass and energy there.

\item Combination of GALFACTS images with those from higher frequency radio surveys and the recent IRIS infra-red images \citep{IRIS} of similar resolution will allow the separation of the thermal and non-thermal components of the large-scale Galactic continuum emission. In particular, derivation of the distribution of the thermal component will provide vital information for the interpretation of the measurements of the GALFA Radio Recombination Line Survey.

\item GALFACTS will cover considerable areas of the Galactic Loops I, II and IV. These large-scale features are usually interpreted as representing emission from nearby, old supernova remnants. These angularly huge, highly polarized objects have yet to be studied with the resolution provided by GALFACTS. Interest in these objects has recently been heightened with the publication of a new interpretation of the emission from Loop I
\citep{Wolleben_2007}.

\item The sensitivity and resolution of GALFACTS offers a unique contribution to the task of removing foreground emission from high-frequency studies of the intensity and polarization of the Cosmic Microwave Background (CMB) radiation. 

\item A search will be made for serendipitous detections of unexpected phenomena including sources of transient emission, and highly circularly polarized radiation from sources such as the 100\% circularly polarized periodic radio bursts recently detected from low-mass stars and Brown Dwarfs \citep{Hallinan_2008}.

\item Carry out a sensitive full-Stokes variability study, which will employ both the Òdouble passÓ nature of the GALFACTS observations for time scales of days and weeks, and comparison with the NVSS catalog for time scales of order 
10 years. 

\end{itemize}

\bibliography{Taylor_Andrew}

\begin{thebibliography}{}
\expandafter\ifx\csname natexlab\endcsname\relax\def\natexlab#1{#1}\fi
\expandafter\ifx\csname url\endcsname\relax
  \def\url#1{\texttt{#1}}\fi
\expandafter\ifx\csname urlprefix\endcsname\relax\def\urlprefix{URL }\fi
\providecommand{\eprint}[2][]{\url{#2}}

\bibitem[{Brown et~al.(2007)Brown, Haverkorn, Gaensler, Taylor, Bizunok,
  McClure-Griffiths, Dickey, \& Green}]{Brown_2007}
Brown, J.~C., Haverkorn, M., Gaensler, B.~M., Taylor, A.~R., Bizunok, N.~S.,
  McClure-Griffiths, N.~M., Dickey, J.~M., \& Green, A.~J. 2007, ApJ, 663, 258

\bibitem[{Desch\^enes \& Lagache(2006)}]{IRIS}
Desch\^enes, M.-A.~M., \& Lagache, G. 2006, in The Spitzer Space Telescope: New
  Views of the Cosmos, edited by L.~Artmus, \& W.~T. Reach (USA: ASP), vol. 357
  of ASP Conference Series., 167

\bibitem[{Grant et~al.(2010)Grant, Taylor, Stil, Landecker, Kothes, Ransom, \&
  Scott}]{Grant_2010}
Grant, J.~K., Taylor, A.~R., Stil, J.~M., Landecker, T.~L., Kothes, R., Ransom,
  R.~R., \& Scott, D. 2010, ApJ, 714, 1689

\bibitem[{Guram(2008)}]{guram_2008}
Guram, S. 2008, M.\ sc. thesis

\bibitem[{Hallinan et~al.(2008)Hallinan, Antonova, Doyle, Bourke, Lane, \&
  Golden}]{Hallinan_2008}
Hallinan, G., Antonova, A., Doyle, J.~G., Bourke, S., Lane, C., \& Golden, A.
  2008, ApJ, 684, 644

\bibitem[{Haverkorn et~al.(2006)Haverkorn, Gaensler, McClure-Griffiths, Dickey,
  \& Green}]{SGPS_pol}
Haverkorn, M., Gaensler, B.~M., McClure-Griffiths, N.~M., Dickey, J.~M., \&
  Green, A.~J. 2006, ApJ, 167, 230

\bibitem[{Haverkorn et~al.(2003)Haverkorn, Katgert, \&
  de~Bruyn}]{Haverkorn_2003}
Haverkorn, M., Katgert, P., \& de~Bruyn, A.~G. 2003, AA, 403, 1045

\bibitem[{Kothes \& Brown(2009)}]{Kothes_2009}
Kothes, R., \& Brown, J.~C. 2009, in Cosmic Magnetic Fields: From Planets, to
  Stars and Galaxies, edited by K.~G. Strassmeier, A.~G. Kosovichev, \& J.~E.
  Beckman (UK: Cambridge University Press), vol. 259 of IAU Symp., 75

\bibitem[{Landecker et~al.(2010)Landecker, Reich, Reid, Reich, Wolleben,
  Kothes, Uyaniker, Gray, Rizzo, F\"urst, Taylor, \& Wielebinski}]{CGPS_pol}
Landecker, T.~L., Reich, W., Reid, R.~I., Reich, P., Wolleben, M., Kothes, R.,
  Uyaniker, B., Gray, A.~D., Rizzo, D.~D., F\"urst, E., Taylor, A.~R., \&
  Wielebinski, R. 2010, AA, 1004, 2536

\bibitem[{Reich et~al.(1997)Reich, Reich, \& F\"urst}]{Reich_1997}
Reich, P., Reich, W., \& F\"urst, E. 1997, A\&A Supp., 126, 413

\bibitem[{Sun et~al.(2007)Sun, Han, Reich, Reich, Shi, Wielebinski, \&
  F\"urst}]{Sun_2007}
Sun, X.~H., Han, J.~L., Reich, W., Reich, P., Shi, W.~B., Wielebinski, R., \&
  F\"urst, E. 2007, AA, 463, 993

\bibitem[{Taylor et~al.(2003)Taylor, Gibson, Peracaula, Martin, Landecker,
  Brunt, Dewdney, Dougherty, Gray, Higgs, Kerton, Knee, Kothes, Purton,
  Uyanker, Wallace, \& Willis}]{CGPS}
Taylor, A.~R., Gibson, J., Peracaula, M., Martin, P.~G., Landecker, T.~L.,
  Brunt, C.~M., Dewdney, P.~E., Dougherty, S.~M., Gray, A.~D., Higgs, L.~A.,
  Kerton, C.~R., Knee, L. B.~G., Kothes, R., Purton, C.~R., Uyanker, B.,
  Wallace, B.~J., \& Willis, A.~G. 2003, AJ, 125, 3145

\bibitem[{Taylor et~al.(2007)Taylor, Stil, Grant, Landecker, Kothes, Reid,
  Gray, Scott, Martin, Boothroyd, Joncas, Lockman, English, Sajina, \&
  Bond}]{Taylor_2007}
Taylor, A.~R., Stil, J.~M., Grant, J.~K., Landecker, T.~L., Kothes, R., Reid,
  R.~I., Gray, A.~D., Scott, D., Martin, P.~G., Boothroyd, A.~I., Joncas, G.,
  Lockman, F.~J., English, J., Sajina, A., \& Bond, J.~R. 2007, ApJ, 666, 201

\bibitem[{Uyaniker et~al.(1999)Uyaniker, F\"urst, Reich, Reich, \&
  Wielebinski}]{Uyaniker_1999}
Uyaniker, B., F\"urst, E., Reich, W., Reich, P., \& Wielebinski, R. 1999,
  A\&AS, 138, 31

\bibitem[{Wolleben(2007)}]{Wolleben_2007}
Wolleben, M. 2007, ApJ, 664, 349

\bibitem[{Wolleben et~al.(2009)Wolleben, Landecker, Carretti, Dickey, Fletcher,
  Gaensler, Han, Haverkorn, Leahy, McClure-Griffiths, McConnel, Reich, \&
  Taylor}]{Wolleben_2009}
Wolleben, M., Landecker, T.~L., Carretti, E., Dickey, J.~M., Fletcher, A.,
  Gaensler, B.~M., Han, J.~L., Haverkorn, M., Leahy, J.~P., McClure-Griffiths,
  N.~M., McConnel, D., Reich, W., \& Taylor, A.~R. 2009, in Cosmic Magnetic
  Fields: From Planets, to Stars and Galaxies, edited by K.~G. Strassmeier,
  A.~G. Kosovichev, \& J.~E. Beckman (UK: Cambridge University Press), vol. 259
  of IAU Symp., 89

\bibitem[{Wolleben et~al.(2006)Wolleben, Landecker, Reich, \&
  Wielebinski}]{Wolleben_2006}
Wolleben, M., Landecker, T.~L., Reich, W., \& Wielebinski, R. 2006, AA, 448,
  411

\end{thebibliography}

\end{document}